\documentclass[preprint,12pt]{elsarticle}
\usepackage{amssymb}
\usepackage{natbib}
\usepackage{subfig}
\usepackage[percent]{overpic} %esto es para para enumerar a las subfiguras: (a),(b),etc.

\begin{document}

%\pacs{89.75.-k, 64.60.aq, 64.60.ah}

\begin{frontmatter}

\title{Crossover from weak to strong disorder regime in the duration
  of epidemics}

\author[mdp]{C. Buono\corref{cor}}
\ead{cbuono@mdp.edu.ar}

\author[mdp]{C. Lagorio}

\author[mdp]{P. A. Macri}

\author[mdp,bst]{L. A. Braunstein}

\cortext[cor]{Corresponding author}

\address[mdp]{Instituto de Investigaciones F\'isicas de Mar del Plata
  (IFIMAR)-Departamento de F\'isica, Facultad de Ciencias Exactas y
  Naturales, Universidad Nacional de Mar del Plata-CONICET, Funes
  3350, (7600) Mar del Plata, Argentina.}

\address[bst]{Center for Polymer Studies, Physics Department, Boston
  University, Boston, Massachusetts 02215, USA.}

\begin{abstract}

We study the Susceptible-Infected-Recovered model in complex networks,
considering that not all individuals in the population interact in the
same way between them. This heterogeneity between contacts is modeled
by a continuous disorder. In our model the disorder represents the
contact time or the closeness between individuals. We find that the
duration time of an epidemic has a crossover with the system size,
from a power law regime to a logarithmic regime depending on the
transmissibility related to the strength of the disorder. Using
percolation theory, we find that the duration of the epidemic scales
as the average length of the branches of the infection. Our
theoretical findings, supported by simulations, explains the
crossover between the two regimes.

\end{abstract}


\end{frontmatter}

\section{Introduction}
\label{sec1}

Complex Networks has became a topic of interest among scientists in
the last years, due to the fact that many real systems such as protein
interaction, Internet, communication systems
\cite{protein,power,internet}, among others, can be properly described
by complex networks, making this theoretical framework inherently
interdisciplinary. On complex networks, nodes can represent the
individuals of a population in the case of social networks, computers
in communication systems, etc, and the links represent the
interaction between them.  The research on networks goes from the
study of its topology to the study of transport processes that use the
networks as the underlying substrate to propagate. In particular, many
researchers have focused on the propagation of seasonal diseases on
social networks due to the appearance of new infections such as SARS,
Avian Flu and the recently influenza A(H1N1) epidemic.

Several models have been developed to characterize the spreading of
this kind of diseases, one of the most used models is the
Susceptible-Infected-Recovered (SIR) model, first introduced by
Kermack and McKendrick \cite{kermack} in the full mixing approximation
and then extended to complex networks \cite{doro,new}.  In this model
the individuals of the population can be in three different states S
(susceptible) - I (Infected) - R (recovered or removed). A S
individual becomes infected with a probability $\beta$ by contact with
an infected individual. Infected individuals recover after $t_r$ times
steps since they were infected, and cannot infect or change their
state thereafter. The system reaches the steady state when all the
infected individuals recover. In this model, the size of the disease
defined as the number of recovered individuals, depends only on the
effective probability of transmission between individuals, given by
\begin{equation}
T= 1-\int(1-\beta)^{t_{r}}P(\beta)d\beta \;,
\label{eq1}
\end{equation}
where $P(\beta)$ is the density distribution of $\beta$, with
$P(\beta)=\delta(\beta - \beta_{0})$ for a constant probability of
infection.

It has been shown that the steady state of the SIR model on statics
complex networks can be mapped into a link percolation problem
\cite{new,gra} where the order parameter of the SIR is the fraction of
recovered individuals, and the control parameter is the
transmissibility $T$ which plays the role of the fraction of links $p$
in percolation. Thus there exist a critical threshold $T_c$ (or $p_c$)
in the SIR (or percolation) model above which a nonzero fraction of
individuals (size of biggest cluster) are infected ($\sim N$).  It was
shown that in finite systems, the critical threshold $T_c\equiv p_c$
of link percolation depends strongly on the network size N with
$T_{c}(N)-T_{c}(N \to \infty)\sim N^{-1/3}$ in mean field
approximation, {\it i.e.} finite size effects are strong
\cite{Wu_01}. Then in finite systems, for $T>T_c(N)$ the disease
becomes an epidemic, while for $T<T_c(N)$ the disease reaches a small
fraction of the population (outbreaks) \cite{new,Mil_02,ken,lago2}.

Usually, the study of a disease transmission or any type of transport
process like information flows \cite{inflow}, rumor spreading
\cite{rumsp} through a network is made assuming that all the contacts
are equivalent. However this assumption does not give a very realistic
description of real networks, such as social networks \cite{actor}
where not all individuals in a society have the same interaction
between them. A way to improve the description of real networks, is to
consider the heterogeneity of the social contacts. This can be done by
considering weighted (disordered) networks, where the weights in the
contacts could represent the closeness or the contact time between the
individuals \cite{bra01,ops,barrat}. The contact time is a parameter
that can be controlled by health policies as a strategy to mitigate
the duration of the spreading of the disease. Using different
strategies like broadcasting, brochures, etc, the public health
agencies can induce people to change their contact time or the
closeness of the contact, for example, encouraging people to reduce
their contact time. This strategy that is a social distancing, was
used already by some governments in the recent wave of influenza
A(H1N1) epidemic in 2009 \cite{ah1n1}. It is known that the disorder
can dramatically alter some topological properties of networks such as
the average length of the optimal paths
\cite{bra01,bra03,scre,porto}. In the optimal path problem defined as
the path between any two nodes that minimizes the total weight along
the path \cite{bra01,bra03,scre}, it was shown that the average length
of the optimal path $l_{opt}$ scales as $N^{\nu_{opt}}$ in the strong
disorder limit (SD), where $\nu _{opt}=1/3$ for homogeneous networks,
and as $l_{opt}\sim \ln N$ in the weak disorder regime (WD), where the
SD limit is related to percolation at criticality
\cite{bra01,bra03,ciep}. However, the exact mapping between the order
parameter of both second order phase transitions, percolation and SIR,
is not affected by the disorder when the disorder is not correlated
\cite{Wu_01}. Nevertheless, the disorder could affect the duration of
a disease.

In this paper we introduce disorder in the links and show how a broad
disorder affects the duration of an epidemic in the SIR model.

Using theoretical arguments, supported by intensive simulations, we
find that, the average time of the duration of the epidemic $t_f$ goes
as the average length of the branches $l_b$ of the infection. Thus,
relating $l_b$ with the optimal path problem we find that $t_f$ has a
crossover from WD to SD regimes.

The paper is organized as follows: In section \ref{sec2} we present
our model, in section \ref{sec3} we show theoretical and by
simulation, the crossover for the duration time of the disease from
WD to SD regimes. Finally, in section \ref{sec4} we present our
conclusions.

\section{SIR model with disorder} 
\label{sec2}

In our model, we assign to each link $i$ between any two nodes a
random number $\beta_{i}$ drawn from the distribution
\begin{equation}
P(\beta_{i})=\frac{1}{a\beta_{i}},
\end{equation}
where $a$ is the parameter that controls the broadness of the
distribution of link weights {\it i.e.} the strength of the disorder
and $\beta_i$ is defined in the interval $[e^{-a},1]$. With this
distribution we randomly assign a weight to each link of the network,
of the form,
\begin{equation} \label{eq.disorder}
\beta_{i}=e^{-a \; r_{i}},
\end{equation}
where $r_{i}$ is a random number taken from an uniform distribution
$r_{i}$ $\epsilon$ $[0,1]$. Thus $\beta_{i}$ is the probability of
infection between any pair of nodes. This type of weight has been
widely used \cite{bra03,ciep,porto,bra02} because it is a well known
example of many distributions that produce WD and SD crossover. The
strong disorder limit can be thought in a disordered medium as a
potential barrier $\epsilon_{i}$ so that $\tau_{i}$ is the time to
cross this barrier in a thermal activation process, then
$\tau_{i}=e^{\epsilon_{i}/k\hat{T}}$, where $k$ is the Boltzmann
constant and $\hat{T}$ is the absolute temperature. Therefore, we can
see $\beta_{i}$ as the inverse of the time needed to cross this
barrier $\tau_{i}$.

With this weight distribution the transmissibility $T$ (See
Eq. \ref{eq1}) is given by,
\begin{equation}
T= 1-\int_{e^{-a}}^{1}\frac{(1-\beta)^{t_{r}}}{a\beta}d\beta \; .
\end{equation}
In our initial stage all the individuals are in the susceptible state.
We choose a node at random from the biggest connected cluster, or
giant component (GC) and infect it (patient zero), then the process
follows the rules of the SIR model but in a weighted network. After
the system reaches the steady state, we compute the duration of the
infection $t^*$ defined as the time where the last infected individual
recovers and the length of the branches of the infection as a function
of $t^{*}$. As the substrate for the disease spreading we use only an
Erd\"os R\'enyi network (ER) \cite{Erd_01}, characterized by a Poisson
degree distribution $P(k)=e^{\langle k\rangle}\langle k \rangle
^{k}k!$ where $k$ is the connectivity and $\langle k \rangle$ is the
average degree. However, the results obtained are similar for all
networks with a finite value of $T_{c}$ in the mean field approach.

\begin{figure}
\centering
\includegraphics[width=70mm]{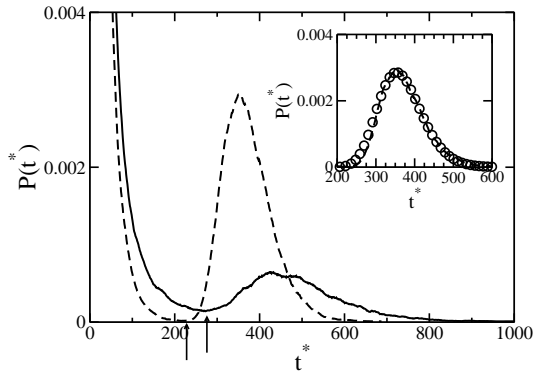}
 \caption{$P(t^{*})$ as a function of $t^{*}$ for an ER network with
   $\langle k\rangle =2$, $N=2^{14}$, $t_{r}=19$, $a=4.36$ (full line)
   and $a=3.24$ (dashed line). The arrows indicate the separation
   between outbreaks (left) and epidemics (right) regime. In the inset
   we show in symbols the fitting with Eq.(\ref{lognorm}) of the data
   for $a=3.24$ (see main plot) from where we obtain $t_f \simeq
   363.8$. All simulations were done over $10^{5}$ realizations.}
 \label{fig1}
\end{figure}

\section{Crossover from WD to SD for the duration times of the epidemics}
\label{sec3}

We only consider those propagation that lead to epidemic states, and
disregarded the outbreaks (See Fig. \ref{fig1}).  Fig. \ref{fig1}
shows $P(t^{*})$ as a function of $t^{*}$, the arrows shows the
separation between outbreaks and epidemic regimes. The criteria used 
to distinguish between outbreaks and epidemics is the observation of 
the behavior of the cluster size distribution that decays as a power 
law close to $T_{c}$ for the outbreaks and has a maximum for the 
epidemics. The same criteria is used in percolation to distinguish 
finite cluster sizes from giant component with size $\sim N$.  We 
find that $P(t^{*})$ in the epidemic regime can be well
represented by a Log-Normal distribution,
\begin{eqnarray}
  P(t^{*})=\frac{1}{t^{*}\sigma \sqrt{2\pi}}
  e^{-\frac{\ln(t^{*}/t_f)^{2}}{2\sigma^{2}}} \; ,
\label{lognorm}
\end{eqnarray}
where $t_{f}$ and $\sigma$ are the average and the standard deviation
of the distribution of $t^*$. Log-normal distributions have been
observed in several phenomena, such as the size of crushed ore
\cite{ore}, fragmentation of glass \cite{glass}, income distribution
\cite{income}, events in medical histories \cite{medical}, food
fragmentation by human mastication \cite{mastication}, etc. By
fitting our data with a log-normal distribution we obtain $t_f$ (see
the inset of Fig. \ref{fig1}).

From Fig. \ref{fig1} we can see that $t_{f}$ increases with $a$ for a
fixed value of $t_{r}$, this behavior can be understood if we take
into account that when $a$ increases, the barrier that the disease
needs to overcome in order to propagate is bigger. Therefore, even
though the transmissibility decreases as $a$ increases for fixed
$t_r$, the epidemic will last longer due to the disorder, allowing the
health services to make earlier interventions.

\begin{figure}
 \centering
 \includegraphics[width=70mm]{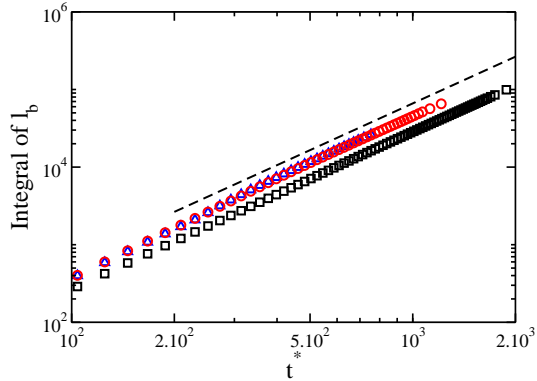}
 \caption{Log-log plot of the integral of $l_b$ as function of $t^*$,
   for an ER network with $\langle k\rangle =2$, $N=2^{14}$ and
   $t_{r}=19$. For the undisordered problem $T=T_{c}$ ($\Box$), and
   for the disordered problem with $T=T_{c}$ and $a=6.55$
   ($\bigcirc$), and $T > T_{c}$ and $a=3.97$ ($\bigtriangleup$). The
   dashed line is used as a guide to show the slope $2$. All
   simulations were done over $10^{5}$ realizations. (Color online)}
 \label{fig4}
\end{figure}

\begin{figure}
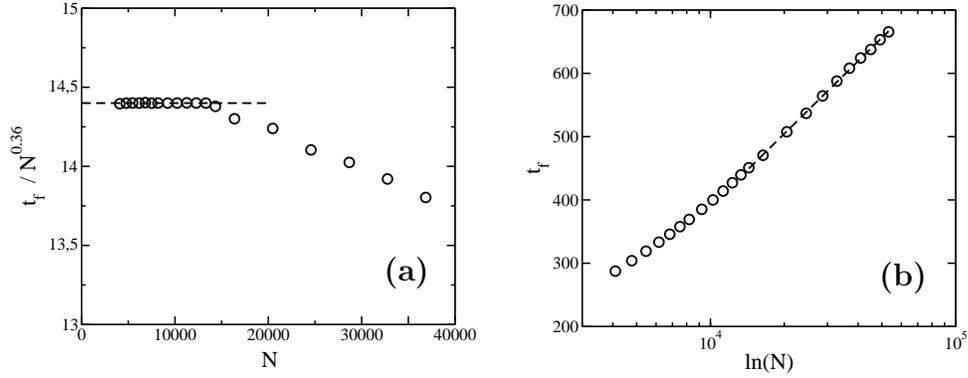

\centering
  \begin{overpic}[scale=0.25]{fig3.eps}
    \put(80,20){{\bf{(a)}}}
  \end{overpic}\hspace{0.5cm}
  \begin{overpic}[scale=0.25]{fig4.eps}
    \put(80,20){\bf{(b)}}
  \end{overpic}\vspace{0.5cm}
 \caption{For an ER network with $\langle k\rangle =2$, $t_{r}=19$ and
   $T=T_{c}(N=2^{12})$. (a) Linear-linear plot of $t_{f}/N^{0.36}$ as
   a function of $N$, the dashed line is used as a guide to show that
   the exponent $0.36$ is correct. (b) Linear-log plot of $t_{f}$ as a
   function of $\ln(N)$, the dashed line correspond to a logarithmic
   fit. Notice that in the WD regime $T_{c}(N=2^{12}) > T_{c}(N>2^{12})$. All
   simulations were done over $10^{5}$ realizations. }
 \label{fig3}
\end{figure}

We also study the average length of the branches of the infection
$l_{b}$. The length of a branch is defined as the number of links
between the patient zero and the last patient in that branch. In
Fig. \ref{fig4} we plot the integral of $l_{b}$ as a function of $t^*$
in log-log scale, from which it is easy to see that, independent of
the disorder, the epidemic region has a slope $2$, thus $l_b\sim t^*$.
If we compare $l_b$ between the disordered and the undisordered
substrate, we can see that in the disordered one, the length of the
branches are larger than in the undisordered case, even for the same
values of $T$ and $t_r$. This is due to the fact that the link of
smallest crossing probability $e^{-a}$ that the disease has to
traverse is smaller than the required value of $\beta$ for the same
values of $T$ and $t_{r}$ in the undisordered case (see
Eq.(\ref{eq1})).

For the SIR problem, at the threshold $T_c$ in the mean field
approximation, the fraction of recovered individuals $R$ grows with
time $t^*$ as $R \sim {t^*}^2$ \cite{kap} up to $t_f$, then at $t_f$,
$R \sim t_f^2$, and for $t^* > t_f$, $R$ $\approx$ constant due to
finite size effects. Percolation in complex networks predicts that at
criticality $T \approx T_c$, the size $R$ of the epidemics scales with
the chemical length $l$ as \cite{Coh_01}.
\begin{eqnarray}
R \sim \ell^{d_{l}} \sim t_{f}^{2} \; ,
\end{eqnarray}
with $d_{l}=2$ in the mean field approximation, then
\begin{eqnarray}
t_f \sim \ell \; ,
\end{eqnarray}
where $\ell \sim N^{1/3}$. For the WD regime where $T > T_c$,
\begin{eqnarray}
 R \sim \left\{ \begin{array}{lcl} e^{t^*} & \mbox{ for } & t^*
   \lesssim t_f \; ; \\ & & \\ const. & \mbox{ for } & t^* > t_f \; ,
\end{array}
\right.
\end{eqnarray}
then at $t_f$
\begin{eqnarray}
R \sim \exp(t_f) \; .
\label{eq9}
\end{eqnarray}
Using the fact that in WD , $R \sim N$ and $\ell \sim \ln N$, we
obtain that also in this regime
\begin{eqnarray}
t_f \sim \ell \; .
\label{tfWD}
\end{eqnarray}
Thus from the theoretical arguments presented above,
\begin{equation}
t_{f} \sim \left\{ 
\begin{array}
{lcl} N^{1/3} & \mbox{if } & T\simeq T_{c} \\ 
& & \\ \ln(N) & \mbox{ if } & T>T_{c}     
\end{array}
\right.
\label{eq11}
\end{equation}

In order to corroborate Eq. (\ref{eq11}), we compute $t_{f}$ as a
function of $N$ for fixed values of $T = T_{c} (N=2^{12})$, $a$ and
$t_{r}$, and found that for $T\simeq T_{c}(N)$, $t_{f}$ behaves as a
power law with exponent $0.36$ compatible with the one found for SD in
the optimal path problem. The slight difference between the exponent
obtained and the theoretical one (1/3) is due to finite size effects
\cite{Wu_01}. Notice that the power law regime of SD holds only until
$N\simeq 15000$. In Fig. \ref{fig3} (a), we plot the ratio
$t_{f}/N^{0.36}$ as a function of $N$. We can see that for values of
$N$ for which $T\simeq T_{c}(N)$, this ratio goes to a constant in
agreement with our prediction. However for $T > T_{c}(N)$ we can see
from Fig. \ref{fig3} (b) that $t_{f}$ behaves logarithmically with $N$
in agreement with the WD regime $T>T_{c}$ of Eq. (\ref{eq11}). Thus,
we can see clearly that there is a crossover from WD to SD in full
agreement with our theoretical prediction.

\section{Conclusions}
\label{sec4}

In this paper we study the SIR model on a broad disordered network,
where the disorder represents the duration of the interactions between
the individuals of a population, and the broadness of the disorder is
a control parameter.

We found that as the broadness of the disorder increases, the
spreading of the disease is delayed. Thus, recommending people to
decrease the duration of the contact as a social distancing could be a
good strategy to delay the spreading of an epidemic. Moreover, this
social distancing strategy, is more suitable than a quarantine, where
contacts are broken, due to the fact that it is less expensive from an
economic point of view.

Using percolation arguments, we found that the duration time of the
epidemic goes as the average length of the branches of the
infection. Our theoretical results are strongly supported by
simulations. Thus, in the same way as in the optimal path problem, the
duration of the disease has a crossover with the system size, from a
power law regime (SD) to a logarithmic regime (WD).

\section*{Acknowledgments}

This work was financially supported by UNMdP and FONCyT (Pict
0293/2008). The authors thank Lucas D. Valdez for useful comments and
discussions.

\bibliographystyle{model1-num-names}
\bibliography{bibpaper-buono.bib}

\begin{thebibliography}{31}
\expandafter\ifx\csname natexlab\endcsname\relax\def\natexlab#1{#1}\fi
\providecommand{\bibinfo}[2]{#2}
\ifx\xfnm\relax \def\xfnm[#1]{\unskip,\space#1}\fi
%Type = Article
\bibitem[{Jeong et~al.(2011)Jeong, Mason, Barab{\'a}si, and Oltvai}]{protein}
\bibinfo{author}{H.~Jeong}, \bibinfo{author}{S.~Mason}, \bibinfo{author}{A.-L.
  Barab{\'a}si}, \bibinfo{author}{Z.~N. Oltvai},
\newblock \bibinfo{journal}{Nature} \bibinfo{volume}{41} (\bibinfo{year}{2011})
  \bibinfo{pages}{411}.
%Type = Article
\bibitem[{Watts and Strogatz(1998)}]{power}
\bibinfo{author}{D.~J. Watts}, \bibinfo{author}{S.~H. Strogatz},
\newblock \bibinfo{journal}{Nature} \bibinfo{volume}{393}
  (\bibinfo{year}{1998}) \bibinfo{pages}{440--442}.
%Type = Book
\bibitem[{Pastor-Satorras and Vespignani(1992)}]{internet}
\bibinfo{author}{R.~Pastor-Satorras}, \bibinfo{author}{A.~Vespignani},
  \bibinfo{title}{Evolution Structure of the internet},
  \bibinfo{publisher}{Oxford University Press, Oxford}, \bibinfo{year}{1992}.
%Type = Article
\bibitem[{Kermack and McKendrick(1927)}]{kermack}
\bibinfo{author}{W.~O. Kermack}, \bibinfo{author}{A.~G. McKendrick},
\newblock \bibinfo{journal}{Proc. Roy. Soc. Lond. A} \bibinfo{volume}{115}
  (\bibinfo{year}{1927}) \bibinfo{pages}{700--721}.
%Type = Book
\bibitem[{Dorogovtsev and Mendes(2003)}]{doro}
\bibinfo{author}{S.~N. Dorogovtsev}, \bibinfo{author}{J.~F.~F. Mendes},
  \bibinfo{title}{Evolution of Networks}, \bibinfo{publisher}{Oxford University
  Press, Oxford}, \bibinfo{year}{2003}.
%Type = Article
\bibitem[{Newman(2002)}]{new}
\bibinfo{author}{M.~E.~J. Newman},
\newblock \bibinfo{journal}{Physical Review E} \bibinfo{volume}{66}
  (\bibinfo{year}{2002}) \bibinfo{pages}{016128}.
%Type = Article
\bibitem[{Grassberger(1983)}]{gra}
\bibinfo{author}{P.~Grassberger},
\newblock \bibinfo{journal}{Math. Biosci.} \bibinfo{volume}{63}
  (\bibinfo{year}{1983}) \bibinfo{pages}{157--172}.
%Type = Article
\bibitem[{Wu et~al.(2007)Wu, Lagorio, Braunstein, Cohen, Havlin, and
  Stanley}]{Wu_01}
\bibinfo{author}{Z.~Wu}, \bibinfo{author}{C.~Lagorio}, \bibinfo{author}{L.~A.
  Braunstein}, \bibinfo{author}{R.~Cohen}, \bibinfo{author}{S.~Havlin},
  \bibinfo{author}{H.~E. Stanley},
\newblock \bibinfo{journal}{Physical Review E} \bibinfo{volume}{75}
  (\bibinfo{year}{2007}) \bibinfo{pages}{066110}.
%Type = Article
\bibitem[{Miller(2007)}]{Mil_02}
\bibinfo{author}{J.~C. Miller},
\newblock \bibinfo{journal}{Phys. Rev. E} \bibinfo{volume}{76}
  (\bibinfo{year}{2007}) \bibinfo{pages}{010101}.
%Type = Article
\bibitem[{Kenah and Robins(2007)}]{ken}
\bibinfo{author}{E.~Kenah}, \bibinfo{author}{J.~M. Robins},
\newblock \bibinfo{journal}{Phys. Rev. E} \bibinfo{volume}{76}
  (\bibinfo{year}{2007}) \bibinfo{pages}{036113}.
%Type = Article
\bibitem[{Lagorio et~al.(2009)Lagorio, Migueles, Braunstein, L{\'o}pez, and
  Macri}]{lago2}
\bibinfo{author}{C.~Lagorio}, \bibinfo{author}{M.~Migueles},
  \bibinfo{author}{L.~Braunstein}, \bibinfo{author}{E.~L{\'o}pez},
  \bibinfo{author}{P.~Macri},
\newblock \bibinfo{journal}{Phys. A} \bibinfo{volume}{388}
  (\bibinfo{year}{2009}) \bibinfo{pages}{755--763}.
%Type = Article
\bibitem[{Helbing et~al.(2006)Helbing, Treiber, and Kestingand}]{inflow}
\bibinfo{author}{D.~Helbing}, \bibinfo{author}{M.~Treiber},
  \bibinfo{author}{A.~Kestingand},
\newblock \bibinfo{journal}{Physica A} \bibinfo{volume}{363}
  (\bibinfo{year}{2006}) \bibinfo{pages}{62--72}.
%Type = Article
\bibitem[{Nekovee et~al.(2007)Nekovee, Moreno, Bianconi, and Marsili}]{rumsp}
\bibinfo{author}{M.~Nekovee}, \bibinfo{author}{Y.~Moreno},
  \bibinfo{author}{G.~Bianconi}, \bibinfo{author}{M.~Marsili},
\newblock \bibinfo{journal}{Physica A} \bibinfo{volume}{374}
  (\bibinfo{year}{2007}) \bibinfo{pages}{457--470}.
%Type = Article
\bibitem[{Barab{\'a}si and Albert(1999)}]{actor}
\bibinfo{author}{A.-L. Barab{\'a}si}, \bibinfo{author}{R.~Albert},
\newblock \bibinfo{journal}{Science} \bibinfo{volume}{286}
  (\bibinfo{year}{1999}) \bibinfo{pages}{509}.
%Type = Article
\bibitem[{Braunstein et~al.(2007)Braunstein, Wu, Chen, Buldyrev, Kalisky,
  Sreenivasan, Cohen, L{\'o}pez, Havlin, and Stanley}]{bra01}
\bibinfo{author}{L.~A. Braunstein}, \bibinfo{author}{Z.~Wu},
  \bibinfo{author}{Y.~Chen}, \bibinfo{author}{S.~V. Buldyrev},
  \bibinfo{author}{T.~Kalisky}, \bibinfo{author}{S.~Sreenivasan},
  \bibinfo{author}{R.~Cohen}, \bibinfo{author}{E.~L{\'o}pez},
  \bibinfo{author}{S.~Havlin}, \bibinfo{author}{H.~E. Stanley},
\newblock \bibinfo{journal}{I. J. Bifurcation and Chaos} \bibinfo{volume}{17}
  (\bibinfo{year}{2007}) \bibinfo{pages}{2215--2255}.
%Type = Article
\bibitem[{Opsahl et~al.(2008)Opsahl, Colizza, Panzarasa, and Ramasco}]{ops}
\bibinfo{author}{T.~Opsahl}, \bibinfo{author}{V.~Colizza},
  \bibinfo{author}{P.~Panzarasa}, \bibinfo{author}{J.~Ramasco},
\newblock \bibinfo{journal}{Phys Rev Lett} \bibinfo{volume}{101}
  (\bibinfo{year}{2008}) \bibinfo{pages}{168702}.
%Type = Article
\bibitem[{Barrat et~al.(2004)Barrat, Barthelemy, Pastor-Satorras, and
  Vespignani}]{barrat}
\bibinfo{author}{A.~Barrat}, \bibinfo{author}{M.~Barthelemy},
  \bibinfo{author}{R.~Pastor-Satorras}, \bibinfo{author}{A.~Vespignani},
\newblock \bibinfo{journal}{Proc. Natl. Acad. Sci. USA} \bibinfo{volume}{101}
  (\bibinfo{year}{2004}) \bibinfo{pages}{3747}.
%Type = Article
\bibitem[{Balcan et~al.(2009)Balcan, Hu, Goncalves, Bajardi, Poletto, Ramasco,
  Paolotti, Perra, Tizzoni, den Broeck, Colizza, , and Vespignani}]{ah1n1}
\bibinfo{author}{D.~Balcan}, \bibinfo{author}{H.~Hu},
  \bibinfo{author}{B.~Goncalves}, \bibinfo{author}{P.~Bajardi},
  \bibinfo{author}{C.~Poletto}, \bibinfo{author}{J.~J. Ramasco},
  \bibinfo{author}{D.~Paolotti}, \bibinfo{author}{N.~Perra},
  \bibinfo{author}{M.~Tizzoni}, \bibinfo{author}{W.~V. den Broeck},
  \bibinfo{author}{V.~Colizza}, , \bibinfo{author}{A.~Vespignani},
\newblock \bibinfo{journal}{BMC Medicine} \bibinfo{volume}{7}
  (\bibinfo{year}{2009}) \bibinfo{pages}{45}.
%Type = Article
\bibitem[{Braunstein et~al.(2003)Braunstein, Buldyrev, Cohen, Havlin, and
  Stanley.}]{bra03}
\bibinfo{author}{L.~A. Braunstein}, \bibinfo{author}{S.~V. Buldyrev},
  \bibinfo{author}{R.~Cohen}, \bibinfo{author}{S.~Havlin},
  \bibinfo{author}{H.~E. Stanley.},
\newblock \bibinfo{journal}{Phys. Rev. Lett.} \bibinfo{volume}{91}
  (\bibinfo{year}{2003}) \bibinfo{pages}{168701}.
%Type = Article
\bibitem[{Sreenivasan et~al.(2004)Sreenivasan, Kalisky, Braunstein, Buldyrev,
  Havlin, and Stanley.}]{scre}
\bibinfo{author}{S.~Sreenivasan}, \bibinfo{author}{T.~Kalisky},
  \bibinfo{author}{L.~Braunstein}, \bibinfo{author}{S.~Buldyrev},
  \bibinfo{author}{S.~Havlin}, \bibinfo{author}{H.~Stanley.},
\newblock \bibinfo{journal}{Phys. Rev. E} \bibinfo{volume}{70}
  (\bibinfo{year}{2004}) \bibinfo{pages}{046133}.
%Type = Article
\bibitem[{Porto et~al.(1998)Porto, Havlin, Roman, and Bunde}]{porto}
\bibinfo{author}{M.~Porto}, \bibinfo{author}{S.~Havlin}, \bibinfo{author}{H.~E.
  Roman}, \bibinfo{author}{A.~Bunde},
\newblock \bibinfo{journal}{Phys. Rev. E} \bibinfo{volume}{58}
  (\bibinfo{year}{1998}) \bibinfo{pages}{5205}.
%Type = Article
\bibitem[{Cieplak et~al.(1996)Cieplak, Maritan, and Banavar}]{ciep}
\bibinfo{author}{M.~Cieplak}, \bibinfo{author}{A.~Maritan},
  \bibinfo{author}{J.~R. Banavar},
\newblock \bibinfo{journal}{Phys. Rev. Lett.} \bibinfo{volume}{76}
  (\bibinfo{year}{1996}) \bibinfo{pages}{3754}.
%Type = Article
\bibitem[{Braunstein et~al.(2001)Braunstein, Buldyrev, Havlin, and
  Stanley}]{bra02}
\bibinfo{author}{L.~A. Braunstein}, \bibinfo{author}{S.~V. Buldyrev},
  \bibinfo{author}{S.~Havlin}, \bibinfo{author}{H.~E. Stanley},
\newblock \bibinfo{journal}{Phys. Rev. E} \bibinfo{volume}{65}
  (\bibinfo{year}{2001}) \bibinfo{pages}{056128}.
%Type = Article
\bibitem[{Erd{\"o}s and R{\'e}nyi(1959)}]{Erd_01}
\bibinfo{author}{P.~Erd{\"o}s}, \bibinfo{author}{A.~R{\'e}nyi},
\newblock \bibinfo{journal}{Publications Mathematicae} \bibinfo{volume}{6}
  (\bibinfo{year}{1959}) \bibinfo{pages}{290--297}.
%Type = Article
\bibitem[{Kolmogorov and Akad(1941)}]{ore}
\bibinfo{author}{A.~N. Kolmogorov}, \bibinfo{author}{D.~Akad},
\newblock \bibinfo{journal}{Nauk SSSR.} \bibinfo{volume}{31}
  (\bibinfo{year}{1941}) \bibinfo{pages}{99--101}.
%Type = Article
\bibitem[{Ishii and Matsushita(1993)}]{glass}
\bibinfo{author}{T.~Ishii}, \bibinfo{author}{M.~Matsushita},
\newblock \bibinfo{journal}{J. Phys. Soc. Jpn.} \bibinfo{volume}{61}
  (\bibinfo{year}{1993}) \bibinfo{pages}{3474--3477}.
%Type = Article
\bibitem[{Montroll and Shlesinger(1983)}]{income}
\bibinfo{author}{E.~W. Montroll}, \bibinfo{author}{M.~F. Shlesinger},
\newblock \bibinfo{journal}{J. Stat. Phys.} \bibinfo{volume}{32}
  (\bibinfo{year}{1983}) \bibinfo{pages}{209--230}.
%Type = Book
\bibitem[{Lawrence(1988)}]{medical}
\bibinfo{author}{R.~J. Lawrence}, \bibinfo{title}{The Lognormal as Event-Time
  Distribution - Theory and Applications}, \bibinfo{publisher}{Inc. Cambridge},
  \bibinfo{year}{1988}.
%Type = Article
\bibitem[{Kobayashi et~al.(2007)Kobayashi, Kohyama, Kobori, Sasaki, and
  Matsushita}]{mastication}
\bibinfo{author}{N.~Kobayashi}, \bibinfo{author}{K.~Kohyama},
  \bibinfo{author}{C.~Kobori}, \bibinfo{author}{Y.~Sasaki},
  \bibinfo{author}{M.~Matsushita},
\newblock \bibinfo{journal}{J. Phys. Soc. Jpn.} \bibinfo{volume}{76}
  (\bibinfo{year}{2007}) \bibinfo{pages}{044002}.
%Type = Article
\bibitem[{Ben-Naim and Krapivsky(2004)}]{kap}
\bibinfo{author}{E.~Ben-Naim}, \bibinfo{author}{P.~L. Krapivsky},
\newblock \bibinfo{journal}{Phys. Rev. E} \bibinfo{volume}{69}
  (\bibinfo{year}{2004}) \bibinfo{pages}{050901}.
%Type = Article
\bibitem[{Cohen et~al.(2000)Cohen, Erez, ben-Avraham ben-Avraham ben-Avraham
  ben-Avraham ben-Avraham ben-Avraham~ben Avraham, and Havlin}]{Coh_01}
\bibinfo{author}{R.~Cohen}, \bibinfo{author}{K.~Erez},
  \bibinfo{author}{D.~ben-Avraham ben-Avraham ben-Avraham ben-Avraham
  ben-Avraham ben-Avraham~ben Avraham}, \bibinfo{author}{S.~Havlin},
\newblock \bibinfo{journal}{Phys. Rev. Lett.} \bibinfo{volume}{85}
  (\bibinfo{year}{2000}) \bibinfo{pages}{4626--4628}.

\end{thebibliography}

\end{document}